# Frustrated Magnetism in FeGe$_3$O$_4$ with a Chiral Trillium Network

*Matt Boswell[1#], Mingyu Xu[1#], Haozhe Wang[1], Mouyang Cheng[2,3,4], N. Li[5], X. F. Sun[5], Haidong Zhou[6], Huibo Cao[7], Mingda Li[2,8], Weiwei Xie[1*]*

1. Department of Chemistry, Michigan State University, East Lansing, MI, 48824 USA
2. Quantum Measurement Group, Massachusetts Institute of Technology, Cambridge, MA, 02139 USA
3. Department of Materials Science and Engineering, Massachusetts Institute of Technology, Cambridge, MA, 02139 USA
4. Center for Computational Science & Engineering, Massachusetts Institute of Technology, Cambridge, MA, 02139 USA
5. Anhui Provincial Key Laboratory of Magnetic Functional Materials and Device, Institutes of Physical Science and Information Technology, Anhui University, Hefei, Anhui 230601, CN
6. Department of Physics and Astronomy, University of Tennessee, Knoxville, TN 37996, USA
7. Neutron Scattering Division, Oak Ridge National Laboratory, Oak Ridge, TN 37831, USA
8. Department of Nuclear Science and Engineering, Massachusetts Institute of Technology, Cambridge, MA, 02139 USA

Corresponding Author: Weiwei Xie (xieweiwe@msu.edu)
#M.B. and M.X. contributed equally.

## *Abstract*

The discovery of new magnetic ground states in geometrically frustrated lattices remains a central challenge in materials science. Here, we report the synthesis, structural characterization, and frustrated magnetic properties of FeGe$_3$O$_4$, a newly identified compound that crystallizes in the noncentrosymmetric cubic space group $P2_13$. In this structure, Fe atoms form an intricate double-trillium lattice with nearest-neighbor Fe-Fe distances of ~4.2 Å, while Ge$^{2+}$ ions mediate magnetic interactions through Fe-Ge-Fe pathways. Field-dependent magnetization at 2 K shows a pronounced nonlinearity, reaching a maximum moment of 2.55(3) μ$_B$/Fe$^{2+}$ at 70 kOe without evidence of saturation. Magnetic susceptibility, heat capacity, and neutron scattering collectively reveal the onset of short-range magnetic interactions near 5 K, with no long-range ordering detected down to 0.06 K. Specific heat measurements demonstrate strong frustration: only ~34% of the expected magnetic entropy is recovered at 2.4 K. Taken together, these results establish FeGe$_3$O$_4$ as a rare example of a geometrically frustrated trillium-lattice magnet, offering a promising platform for exploring exotic quantum magnetic phenomena.

# Introduction

Geometrically frustrated magnets are of significant interest due to their macroscopic ground-state degeneracies, which give rise to a variety of exotic quantum effects.[1–3] A common feature in the interaction networks of frustrated materials is the presence of odd-membered rings, such as triangles and pentagons.[4] Despite the diverse structural topologies that can give rise to frustration, much of the research in this field has focused on a relatively limited subset of structure types commonly found in oxide materials, including pyrochlore, triangular, Kagome, and face-centered cubic lattices.[5–7] Expanding the study of geometric frustration to less common lattice geometry offers the potential to uncover novel magnetic phases and their associated physics. Recently, the cubic trillium lattice, characterized by the chiral space group $P2_13$, has been investigated as a framework for frustrated magnetism. This lattice features a network of corner-sharing triangles, which naturally predisposes it to geometric frustration.[8–12] Although magnets adopting the trillium lattice are rare, primarily intermetallics referred to as B20 compounds, they exhibit a wide array of intriguing physical phenomena.[13] For instance, in MnSi and CoSi, magnetic skyrmions emerge due to the interplay between ferromagnetism and the Dzyaloshinskii-Moriya interaction.[14,15] FeSi has been proposed as a $d$-electron topological Kondo insulator candidate.[16] Other compounds, such as EuPtSi and EuPtGe, display behavior consistent with strongly correlated spin liquids,[17,18] while CeIrSi is considered a candidate for Ising trillium spin ice.[19] Theoretically, a simple nearest-neighbor Heisenberg antiferromagnet on the chiral trillium lattice has been predicted to support a classical spin-liquid (CSL) phase over a wide temperature range, with 120° helical order in its frustrated magnetic ground state.[11,20] From a structural perspective, B20 compounds such as MSi (M = Mn, Fe, Co) contain a single trillium lattice with nearest M-M distances around 2.7 Å. These short distances often promote long-range magnetic ordering, even under conditions of geometric frustration.[21] For instance, external stimuli, such as pressure-induced quantum phase transitions, can drive emergent behaviors in these systems.[22,23] However, theoretical studies suggest that the degree of frustration in single trillium lattices like those in B20 compounds is insufficient to suppress long-range order entirely due to the short M-M distance.

Later investigations, both experimental and computational, have revealed that quantum spin liquid (QSL) behavior can arise in systems with interconnected trillium lattices. Representative examples include compounds in the langbeinite family, $K_2M_2(SO_4)_3$ (M = Fe, Co,

Mn, Cr), where the interconnected trillium lattices feature nearest M-M distances ranging from 4.4 to 6.2 Å.[24,25] Another notable system, NaMn(HCOO)$_3$, contains magnetic Mn$^{2+}$ ($S$ = 5/2) ions arranged in a trillium lattice with nearest-neighbor Mn-Mn distances of approximately 5.6 Å.[26] The compound K$_2$M$_2$(SO$_4$)$_3$ features a network of trigonally distorted MO$_6$ octahedra, which are interconnected via SO$_4^{2-}$ groups. This connectivity establishes an M-O-S-O-M super-super-exchange pathway that mediates magnetic interactions between M$^{2+}$ ions. Within this structure, there are two distinct crystallographic M$^{2+}$ sites, differentiated by their M-O bond distances, with each site forming a single trillium lattice. Similarly, in NaMn(HCOO)$_3$, neighboring Mn$^{2+}$ cations are connected by HCOO$^-$ ions, resulting in an Mn-O-C-O-Mn superexchange pathway. Both compounds exhibit characters of geometric frustration, including the suppression of long-range magnetic order and the emergence of spin-liquid-like behavior, driven by the intricate interplay of their structural and magnetic interactions.

To the best of our knowledge, no compound has been reported to adopt the chiral trillium network with significantly larger M-M distances that would effectively suppress long-range magnetic interactions via direct exchange. Moreover, such a system featuring an M-O-M or M-T-M pathway for magnetic super-exchange has not yet been observed. The lack of compounds with these characteristics represents a critical gap in the exploration of geometric frustration and the potential emergence of novel magnetic phases in the trillium lattice framework. On the other hand, Fe and Ge form various binary phases, including the B20 FeGe phase.[27] Based on this, our research focused on Fe-Ge-O ternary phases to investigate the potential for discovering novel trillium phases. Herein, we present a comprehensive structural characterization, along with an analysis of the magnetic frustration of FeGe$_3$O$_4$, emphasizing the Fe-Fe exchange interactions within the trillium lattice.

**Experiments and Calculations**

**Synthesis of FeGe₃O₄ Crystals:** FeGe$_3$O$_4$ crystals were synthesized using the chemical vapor transport (CVT) method. Stoichiometric amounts of Fe granules (Alfa Aesar, 99.98%), Fe$_2$O$_3$ powder (JT Baker, Baker Analyzed Reagent Grade), finely ground Ge pieces (Thermo Scientific, 99.9999+%), and GeO$_2$ powder (Alfa Aesar, 99.999%) were mixed thoroughly in an atomic ratio of 2:2:9:9, with a total mass of approximately 300 mg. To facilitate the transport process, ~ 50 mg I$_2$ flakes (Fisher Chemical) were added as the chemical transport agent. The mixture was sealed in a quartz tube under vacuum (~10$^{-5}$ torr) and subjected to a thermal gradient by heating to 600 °C for one week. Subsequently, the tubes were cooled to room temperature at a controlled rate of 10 °C per hour. As a result, red transparent crystals with dimensions of approximately 1-2 mm were successfully grown. All products are stable toward decomposition in air and moisture.

**Phase Analyses and Chemical Compositions:** The synthesized samples were finely ground and analyzed for phase identification and purity using powder X-ray diffraction (PXRD) on a Bruker Davinci powder X-ray diffractometer equipped with Cu Kα radiation ($\lambda_{K\alpha}$ = 1.5406 Å). The upper and lower discriminator values were set to 0.40 V and 0.18 V, respectively, to mitigate the background due to fluorescence. Diffraction patterns were collected over a 2θ range of 5°-120° with a step size of 0.010° in step-scan mode, utilizing a scintillation detector. Phase identification and lattice parameter determination were performed through Rietveld refinements using the GSAS-II software package.[28] Chemical composition analysis was conducted using a JEOL 6610LV scanning electron microscope (SEM) coupled with an energy-dispersive X-ray spectroscopy (EDS) detector (Oxford Instruments Isis X-ray analyzer). Samples were affixed to carbon tape before placement in the SEM chamber and analyzed at an accelerating voltage of 20 kV. Spectra were acquired with a collection time of 100 s, examining multiple points within each phase across various grains. Compositional estimates were refined by Oxford's SEM Quant software, incorporating corrections for matrix effects to ensure accuracy.

**Crystal Structure Determination:** A single crystal with dimensions of 0.098 × 0.063 × 0.029 mm$^3$ was picked up, mounted on a nylon loop with paratone oil, and measured using a Rigaku XtalLAB Synergy, Dualflex, Hypix single crystal X-ray diffractometer equipped with an Oxford Cryosystems 800 low-temperature device. Data acquisition was performed using $\omega$ scans with Mo $K_\alpha$ radiation ($\lambda$ = 0.71073 Å, micro-focus sealed X-ray tube, 50 kV, 1 mA). The measurement strategy, including the total number of runs and images, was determined using the strategy calculation feature in CrysAlisPro software (version 1.171.43.143a, Rigaku OD, 2024). Data reduction induced correction for Lorentz polarization. Numerical absorption correction based on Gaussian integration over a multifaceted crystal model. Empirical absorption correction was applied using spherical harmonics implemented in SCALE3 ABSPACK scaling algorithm. Structure solutions and refinement were conducted using SHELXTL Software Package.[29]

**Physical Properties Measurements of FeGe$_3$O$_4$:** Temperature- and magnetic-field-dependent magnetization measurements of FeGe$_3$O$_4$ single crystals were carried out using a Quantum Design, Magnetic Property Measurement System (MPMS3). The direct-current magnetic susceptibility measurements were performed at the temperature range of 2-300 K under both zero-field-cooled (ZFC) and field-cooled (FC) modes, with an applied field of 100 Oe and 1 kOe. Field-dependent isothermal magnetization measurements were also performed employing magnetic fields ranging from 0 to 7 T and at various temperatures. Temperature-dependent specific heat measurements on the crystals ranging in mass from 1 to 3 mg were carried out using a Quantum Design, Physical Property Measurement System (PPMS DynaCool) in the temperature range of 2.4-100 K at various applied magnetic fields up to 9 T. Low temperature specific heat measurements on the crystals with the total mass ~6.4 mg were carried out down to 0.06 K without an applied magnetic field.

**Single Crystal Neutron Diffraction:** Single crystal neutron diffraction was conducted at CORELLI at the Spallation Neutron Source at Oak Ridge National Laboratory. Two crystals were measured both at 2 K and 20 K with angles ranging from 0° to 360° incrementing 3 degrees.

## Results and Discussions

**Synthesis and Structural Analysis of FeGe$_3$O$_4$:** Synthetic attempts to prepare FeGe$_3$O$_4$ by reacting Fe$_2$O$_3$, GeO$_2$, and Ge as a reducing agent resulted in mixtures of FeGe$_3$O$_4$ and FeGeO$_3$ phases, as predicted by the phase diagram in **Fig. S1**. To gain a deeper insight into the structural features of FeGe$_3$O$_4$, single-crystal X-ray diffraction analysis was performed, focusing on elemental distributions, interatomic distances, and coordination environments, shown in **Fig. 1a-d**. The results of single-crystal diffraction are detailed in **Tables S1** and **S2.** FeGe$_3$O$_4$ crystallizes in the noncentrosymmetric cubic space group $P2_13$ (No. 198), with two distinct Fe sites occupying the 4a Wyckoff positions. Although Fe$^{2+}$ and Ge$^{2+}$ were initially considered as mixed occupancies at these sites, refinement indicated no observable site mixing. The Fe atoms in the 4a sites form a unique double trillium lattice. This interconnected lattice structure exhibits nearest Fe-Fe distances of approximately 4.2 Å.

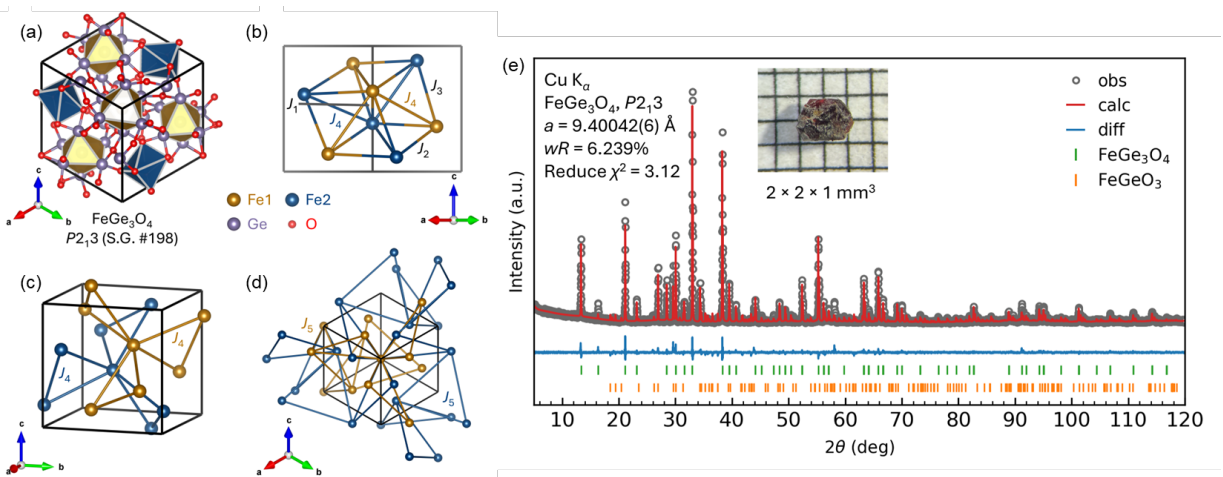

**Fig. 1.** (*a*) Crystal structure of FeGe$_3$O$_4$ ($P2_13$, S.G. 198) with two distinct [FeGe$_6$] and [FeO$_6$] polyhedra shown. Brown, blue, violet, and red represent Fe1, Fe2, Ge, and O atoms. (*b-d*) Exchange interactions ($J_1$ to $J_5$ between Fe$^{2+}$ ions). (*e*) Powder XRD pattern and Rietveld refinement of FeGe$_3$O$_4$. Bragg peak positions of each phase included are represented by vertical tick marks. FeGeO$_3$ exists as impurity powder. (Inset) Optical microscope image (1 mm graph paper) of FeGe$_3$O$_4$.

For the powder X-ray diffraction (PXRD) patterns, all scale factors and lattice parameters were refined, resulting in a reduced $\chi^2$ value of approximately 3.12 and weighted profile residuals ($R_{wp}$) of 6.2%. The diffraction pattern shown in **Fig. 1e** confirms that FeGe$_3$O$_4$ is the predominant phase in the synthesized products, with a minor secondary phase of FeGeO$_3$ present at an estimated concentration of less than 5%. Consistent with single-crystal diffraction and SEM analyses, all refinements confirm a molar ratio of Fe:Ge close to 1:3 for this phase. This agreement across multiple characterization techniques validates the structural and compositional integrity of the synthesized FeGe$_3$O$_4$ sample. Single crystals of FeGe$_3$O$_4$ were manually selected for physical property measurements to ensure sample quality and phase purity.

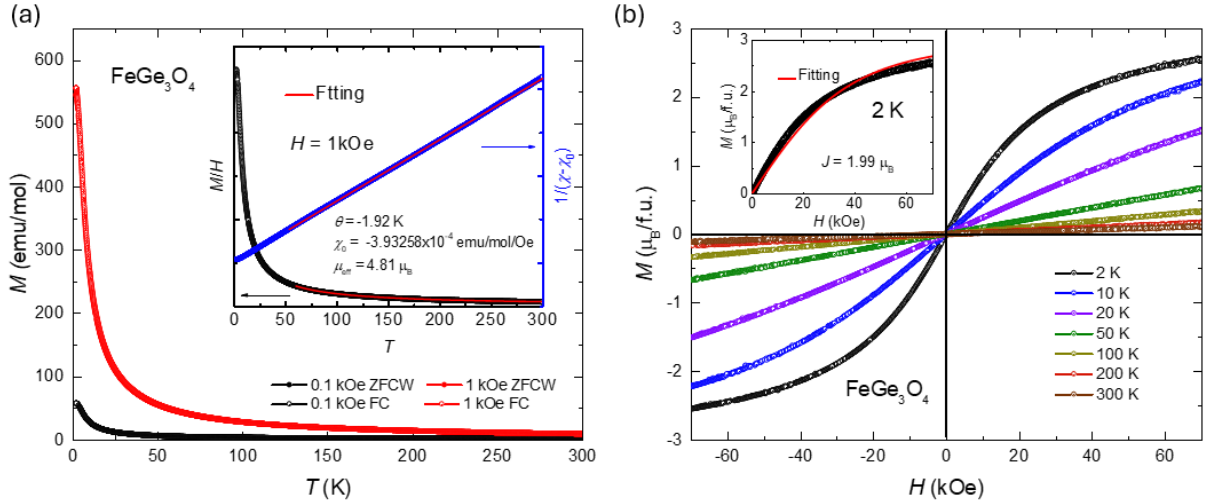

**Fig. 2** (*a*) Temperature dependence of magnetic susceptibility at 0.1 and 1 kOe. Inset, the Curie–Weiss fitting result. (*b*) Magnetic field dependence of magnetization with Brillouin function fitting.

The temperature dependence of the magnetization in both ZFC and FC modes is presented in **Fig. 2a** with no significant difference. The temperature-dependent magnetization measurement shows tail-like behavior without any feature indicating phase transition. As shown in the inset of **Fig. 2a**, magnetization at 1 kOe is fitted using Curie–Weiss (CW) law ($\chi = C/(T - \theta) + \chi_0$) from 60 K to 300 K. Here, $\chi$ represents the magnetic susceptibility, C is the Curie constant, T is the temperature, $\theta$ is the Weiss temperature, and $\chi_0$ is a temperature-independent susceptibility term. The effective moment is around 4.81 $\mu_B$, which suggests that the Fe ion's valance is +2 and is at a high spin state (high-spin Fe$^{2+}$ $\mu_{eff}$ = 4.9 $\mu_B$). This result consists of the hypothesis of the FeGe$_3$O$_4$ valance states in the phase diagram shown in **Fig. S1**. Also, the Weiss temperature is negative (-2 K), which indicates dominant antiferromagnetic interactions in the FeGe$_3$O$_4$, and is similar to the one observed in another quantum spin liquid candidate YbMgGaO$_4$ ($\theta$ = -4 K).[30]

As shown in **Fig. 2b**, field-dependent magnetization measurements show nonlinear behavior at 2 K. The observed maximum magnetization, $M_{max}$ = 2.55(3) $\mu_B$/Fe, was not saturated when the maximum field of 70 kOe was applied. Magnetization curves measured at various temperatures exhibit field-dependent behavior that can be fitted using the Brillouin function as follows[31],

$$M = M_s B_J \left(\frac{g_J \mu_B J B}{k_B T}\right),$$

$$M_s = g_J J,$$

$$g_J = \frac{3J(J+1) + S(S+1) - L(L+1)}{2J(J+1)},$$

$$B_J\left(\frac{g_J \mu_B J B}{k_B T}\right) = \frac{2J+1}{2J}\coth\left(\left(\frac{2J+1}{2J}\right)\left(\frac{g_J \mu_B J B}{k_B T}\right)\right) - \frac{1}{2J}\coth\left(\left(\frac{1}{2J}\right)\left(\frac{g_J \mu_B J B}{k_B T}\right)\right).$$

$M$ is the magnetic moment; $M_s$ is saturation magnetization; $g_J$ is Landé g-factor; $B_J$ is the Brillouin function. Since the CW fitting result agrees with the high-spin state of $Fe^{2+}$, The $S$ and $L$ should be 2. As shown in the inset of **Fig. 2b**, the total moment from the Brillouin function fitting, $J$, is 2 and is in the range of $|L-S|$ and $L+S$. The deviation in the fit may arise from how accurately the temperature captures the effects of thermal frustration. Fitting is much better as a free temperature parameter at around 3 K instead of fixing at 2 K.

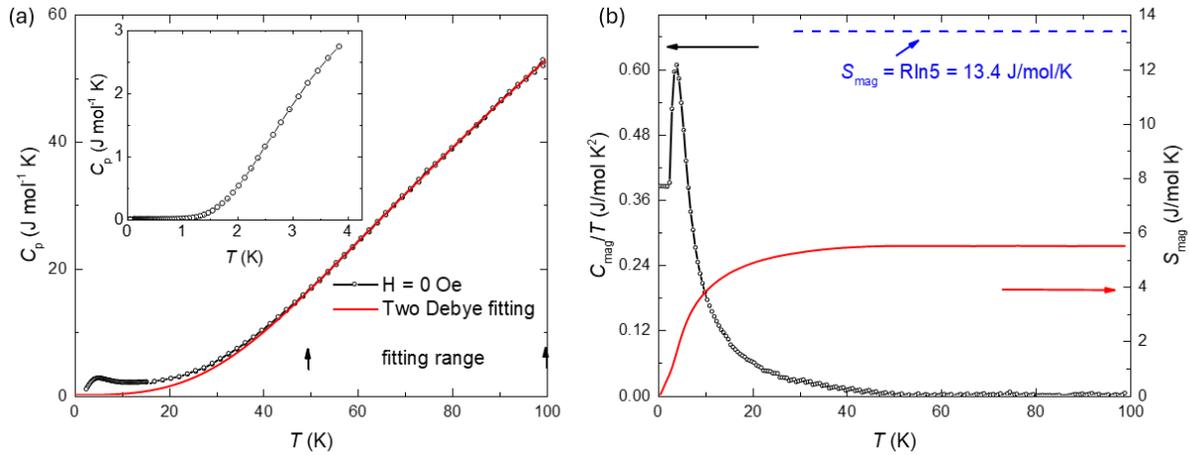

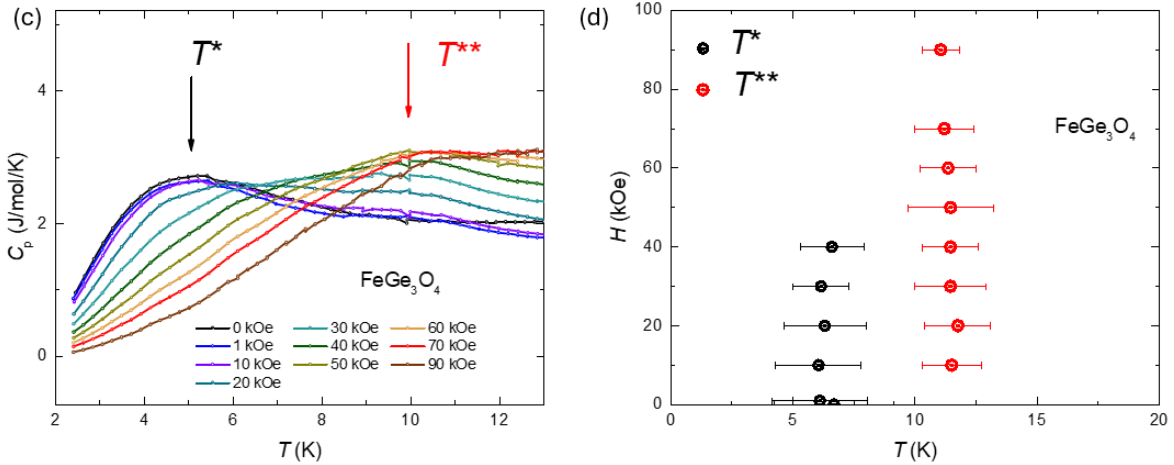

**Fig. 3** (*a*) Temperature-dependent total specific heat of FeGe$_3$O$_4$ at the zero field, together with the two Debye models representing the phonon contribution. The inset gives the low-temperature specific heat measurement. (*b*) Zero field temperature dependence of the magnetic specific heat and magnetic entropy. (*c*) Temperature-dependent specific heat for various magnetic fields. $T^*$ and $T^{**}$ give the two peaks in the temperature-dependent specific heat. (*d*) Temperature-magnetic field phase diagram of FeGe$_3$O$_4$.

**Fig. 3a** shows the temperature-dependent specific heat of FeGe$_3$O$_4$. The transition feature appears around 10 K. The phonon contribution exists dominantly in FeGe$_3$O$_4$ when the temperatures are above 40 K. If we consider only two contributions in the specific heat, $C_p = C_{\text{phonon}} + C_{\text{mag}}$, the data was fitted using two Debye model with the temperature ranging from 40 to 100 K. This yields $\theta_{D1}$ = 307(3) K and $\theta_{D2}$ = 887(13) K, shown in **Fig. 3a**. The inset gives the low-temperature specific heat measurement and there is no transition observed before 0.061 K. The magnetic specific heat Cm is obtained by subtracting the phonon contribution from the fitting. The total entropy $S_{\text{mag}} = \int(C_m/T)dT$ was calculated to be around 4.6 J/mol-K. The magnetic entropy should be $S_{\text{mag}} = R\ln(2J+1)$. Since $J$ =2, as shown in **Fig. 3b**, the saturation magnetic entropy is Rln(5) = 13.4 J/mol/K, and only 34% of the entropy is detected above 2 K. This indicates that FeGe$_3$O$_4$ should be the quantum spin liquid and there is no long-range order detected, which is not similar to other trillium compounds, such as K$_2$Ni$_2$(SO$_4$)$_3$.[32] As shown in **Fig. 3c**, the specific heat exhibits evolving features around 10 K with increasing magnetic field, though no sharp peaks are observed. Given the absence of anomalies in the temperature-dependent magnetization data between 2 K and 300 K, we attribute these specific heat features to the short-range magnetic order. Below 40 K, the temperature-dependent specific heat displays a broad maximum near 5 K at zero magnetic field. As the magnetic field increases, this broad peak (T* ~ 6.5 K) diminishes, while another broader maximum (T** ~11.3 K) emerges and grows more prominent, as shown in the temperature-magnetic field phase diagram **Fig. 3d**.

The lack of magnetic ordering is further confirmed with elastic neutron scattering. Single crystal neutron scattering was conducted at Oak Ridge National Lab beamline 9, CORELLI. A comparison between 2 K and 20 K single crystal neutron scattering is shown in **Fig. 4**. Between 2 K and 20 K the intensities in the low Q range match identically within error, which is further seen from a direct subtraction (**Figs. 4cd**). The lack of any order seen in neutron scattering further confirms the lack of ordering seen in the previously mentioned measurements.

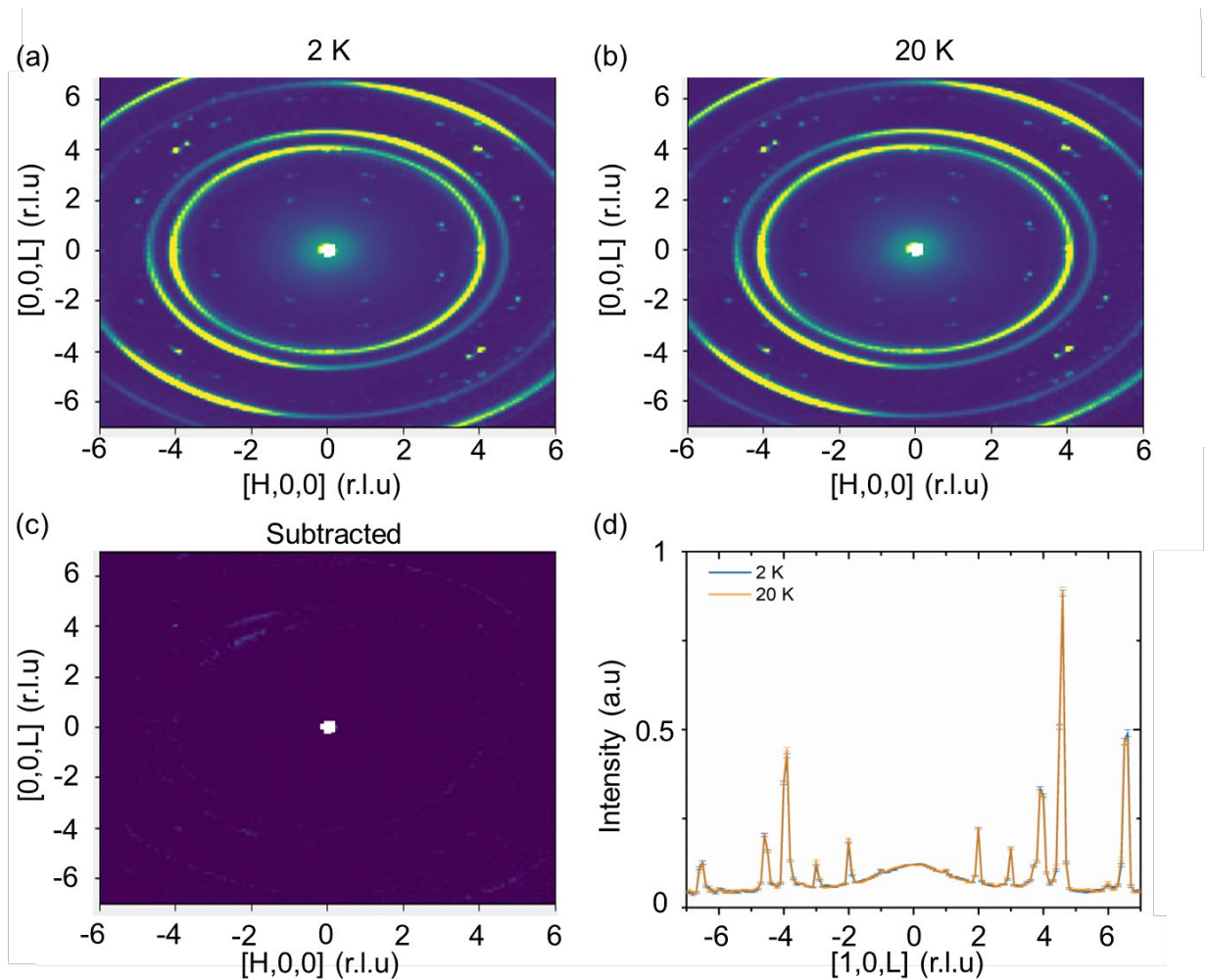

**Fig. 4.** Single crystal neutron diffraction map of FeGe$_3$O$_4$. (HK0) plane plots at (***a***) 2 K and (***b***) 20 K. (***c***) Directly subtracted map highlighting the difference between 2 K and 20 K. (***d***) Comparative intensity plots of (10L) cut at the two temperatures.

## Conclusion

In this work, we report the synthesis and comprehensive characterization of FeGe$_3$O$_4$, a previously unrecognized oxide that crystallizes in the noncentrosymmetric cubic space group $P2_13$ and hosts an intricate double-trillium lattice of Fe atoms. Single-crystal structural refinements confirm well-defined, fully ordered Fe sites with no detectable site mixing, establishing a robust structural platform for unconventional magnetism. Magnetic susceptibility and heat-capacity measurements reveal no signatures of long-range magnetic order down to 0.06 K, despite the presence of strong local moments, which is an immediate indication of geometric frustration within the Fe sublattice. The nonlinear but non-saturated magnetization at low temperature, together with a broad specific-heat anomaly, substantial suppression of magnetic entropy, and diffuse features in neutron scattering, all point toward a short-range correlated, highly frustrated magnetic state. These combined experimental results establish FeGe$_3$O$_4$ as a rare oxide platform in which geometric frustration, competing exchange interactions, and noncentrosymmetric lattice symmetry converge, offering fertile ground for exploring unconventional magnetic phenomena in three-dimensional frustrated systems.

## Supporting Information

This material is available free of charge.

Crystal data and structure refinement of FeGe$_3$O$_4$ at 80 K; Atomic coordinates and isotropic atomic displacement parameters (Å$^2$); Cation's information of Fe-Ge-O systems; Phase Diagram of Fe-Ge-O; Magnetic susceptibility and magnetic field dependence of magnetization; Total and Decomposition of band structure for FeGe$_3$O$_4$; DFT phonon band structure and total density of states for FeGe$_3$O$_4$; Partial phonon density of states for FeGe$_3$O$_4$; Magnetic exchange energies.

## Acknowledgement

The work at Michigan State University was supported by the U.S.DOE-BES under Contract DE-SC0023648. H.Z. is thankful for the support from National Science Foundation Grant No. DMR-2003117. M.L. acknowledges support from U.S. DOE-BES under Contract DE-SC0020148. This research at ORNL's Spallation Neutron Source was sponsored by the Scientific User Facilities Division, Office of Basic Energy Sciences, US Department of Energy. The beamline CORELLI was used through IPTS-36066. This material is based upon work supported by the U.S. Department of Energy, Office of Science, Office of Workforce Development for Teachers and Scientists, Office of Science Graduate Student Research (SCGSR) program. The SCGSR program is administered by the Oak Ridge Institute for Science and Education (ORISE) for the DOE. ORISE is managed by ORAU under contract number DE-SC0014664. All opinions expressed in this paper are the author's and do not necessarily reflect the policies and views of DOE, ORAU, or ORISE. N.L. and X.F.S. were supported by the National Key Research and Development Program of China (Grant No. 2023YFA1406500) and the National Natural Science Foundation of China (Grant Nos. 12404043, 12574042 and 12274388)

**Table of Contents Only**

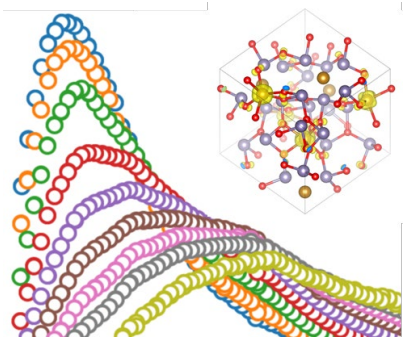

The synthesis, structural characterization, and magnetic frustrated behaviors of FeGe$_3$O$_4$, a newly discovered compound crystallizing in the noncentrosymmetric cubic space group P2$_1$3, were presented.

# Supporting Information

# Frustrated Magnetism in FeGe$_3$O$_4$ with a Chiral Trillium Network

*Matt Boswell[1#], Mingyu Xu[1#], Haozhe Wang[1], Mouyang Cheng[2,3,4], N. Li[5], X.F. Sun[5], Haidong Zhou[6], Huibo Cao[7], Mingda Li[2,8], Weiwei Xie[1*]*

9. Department of Chemistry, Michigan State University, East Lansing, MI, 48824 USA
10. Quantum Measurement Group, Massachusetts Institute of Technology, Cambridge, MA, 02139 USA
11. Department of Materials Science and Engineering, Massachusetts Institute of Technology, Cambridge, MA, 02139 USA
12. Center for Computational Science & Engineering, Massachusetts Institute of Technology, Cambridge, MA, 02139 USA
13. Anhui Provincial Key Laboratory of Magnetic Functional Materials and Device, Institutes of Physical Science and Information Technology, Anhui University, Hefei, Anhui 230601, CN
14. Department of Physics and Astronomy, University of Tennessee, Knoxville, TN 37996, USA
15. Neutron Scattering Division, Oak Ridge National Laboratory, Oak Ridge, TN 37831, USA
16. Department of Nuclear Science and Engineering, Massachusetts Institute of Technology, Cambridge, MA, 02139 USA

Corresponding Author: Weiwei Xie (xieweiwe@msu.edu)
#M.B. and M.X. contributed equally.

## Table of Content

**Table S1.** Crystal data and structure refinement of FeGe$_3$O$_4$ at 80 K.

| Chemical formula | FeGe$_3$O$_4$ |
|---|---|
| Temperature | 80(2) K |
| Formula weight | 337.62 g/mol |
| Space group | P2$_1$3 |
| Unit cell dimensions | a = 9.39775(5) Å |
| Volume | 829.987(14) Å$^3$ |
| Z | 8 |
| Density (calculated) | 5.404 g/cm$^3$ |
| Absorption coefficient | 24.829 mm$^{-1}$ |
| F(000) | 1232 |
| θ range | 3.065 to 41.082° |
| Reflections collected | 53582 |
| Independent reflections | 1844 [$R_{int}$ = 0.0879] |
| Refinement method | Full-matrix least-squares on $F^2$ |
| Data / restraints / parameters | 1844 / 0 / 50 |
| Final R indices | $R_1$ (I>2σ(I)) = 0.0138; $wR_2$ (I>2σ(I)) = 0.0323 |
| | $R_1$ (all) = 0.0146; $wR_2$ (all) = 0.0325 |
| Largest diff. peak and hole | +0.835 e/Å$^3$ and −0.831 e/Å$^3$ |
| R.M.S. deviation from mean | 0.142 e/Å$^3$ |
| Goodness-of-fit on $F^2$ | 1.140 |

**Table S2.** Atomic coordinates and equivalent isotropic atomic displacement parameters (Å$^2$) of FeGe$_3$O$_4$ at 80 K. $U_{eq}$ is defined as one third of the trace of the orthogonalized $U_{ij}$ tensor.

|  | Wyck. | x | y | z | Occ. | $U_{eq}$ |
|---|---|---|---|---|---|---|
| Ge1 | 12b | 0.85422(2) | 0.61460(2) | 0.89612(2) | 1 | 0.00231(4) |
| Ge2 | 12b | 0.10040(2) | 0.86920(2) | 0.85147(2) | 1 | 0.00204(4) |
| Fe1 | 4a | 0.85963(2) | 0.85963(2) | 0.85963(2) | 1 | 0.00178(7) |
| Fe2 | 4a | 0.61569(3) | 0.38431(3) | 0.88431(3) | 1 | 0.00260(7) |
| O1 | 12b | 0.80653(15) | 0.46282(15) | 0.78776(15) | 1 | 0.0042(2) |
| O2 | 12b | 0.70131(14) | 0.55996(15) | 0.00343(14) | 1 | 0.0042(2) |
| O3 | 4a | 0.19443(14) | 0.69443(14) | 0.80557(14) | 1 | 0.0033(3) |
| O4 | 4a | 0.01427(15) | 0.51427(15) | 0.98573(15) | 1 | 0.0039(3) |

**Table S3.** Information on cation compositions and cation valence in the Fe-Ge-O.

| Compounds | Space Group | Fe | Ge | O |
|---|---|---|---|---|
| $Fe_3Ge_2O_8$ | $P2_1/c$ | 2+/3+ | 4+ | -2 |
| $Fe_{15}Ge_8O_{36}$ | P-1 | 2+/3+ | 4+ | -2 |
| $FeGeO_3$ | C2/c | 2+ | 4+ | -2 |
| $Fe_2GeO_4$ | Fd-3m | 2+ | 4+ | -2 |
| $FeGe_3O_4$ | $P2_13$ | 2+ | 2+ | -2 |

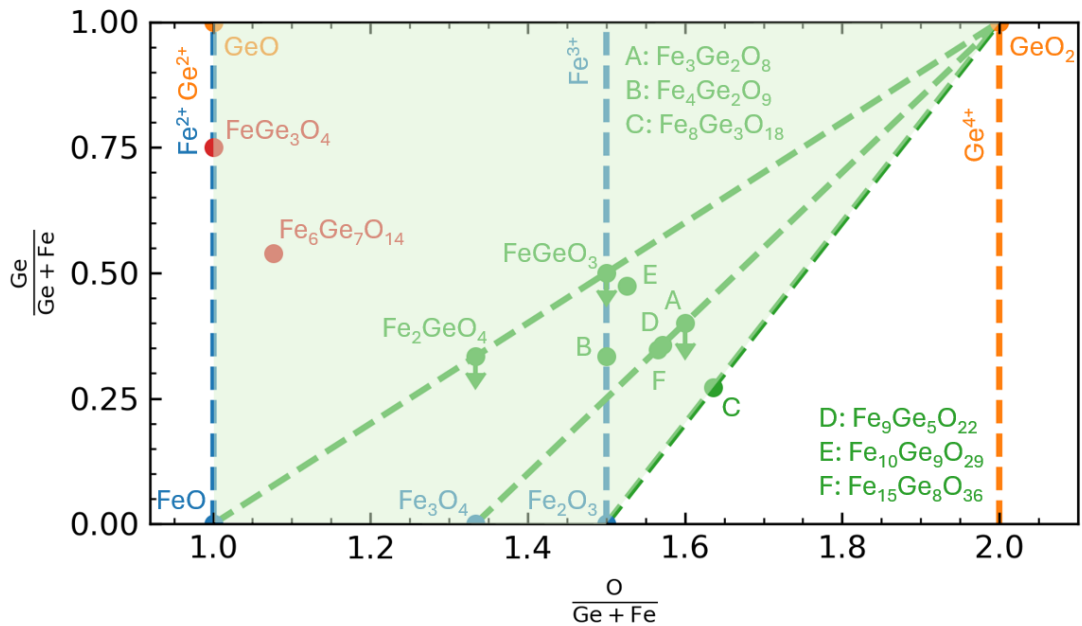

**Fig. S1.** Chemical composition phase diagram of Fe-Ge-O ternary system. Data points represent some of the reported Fe-Ge-O phases. Arrows indicate chemical doping (Fe or Ge) of the parent phases. The green shadow suggests the area of existing Fe-Ge-O ternary compounds.

**Chemical Composition Mapping for New Oxides:** Before delving into the structural and property details of FeGe$_3$O$_4$, we introduce a novel chemical composition map for empirically exploring new oxide materials, as illustrated in **Fig. S1**. The map utilizes the *x*-axis to represent the O/(Ge+Fe) ratio. For reference, the O/(Ge+Fe) ratio is 2 in GeO$_2$ and 1 in GeO, while the O/(Ge+Fe) ratios are 1.5 in Fe$_2$O$_3$, 1.33 in Fe$_3$O$_4$, and 1 in FeO. Fe typically exhibits stable oxidation states of 2+ and 3+, whereas Ge stabilizes in oxidation states of 2+ and 4+. Under mild experimental conditions, ternary Fe-Ge-O phases generally incorporate combinations of Fe$^{3+}$, Fe$^{2+}$, Ge$^{2+}$, and Ge$^{4+}$. This allows us to construct compositional lines connecting GeO$_2$ with Fe$_2$O$_3$, Fe$_3$O$_4$, and FeO. As anticipated, most of the reported ternary phases (highlighted in green) fall along these dashed lines. Theoretical constraints suggest that ternary phases cannot exist beyond the boundary defined by the GeO(blue)- Fe$_2$O$_3$(green) line due to the maximum oxidation states of Ge$^{4+}$ and Fe$^{3+}$ (The green shadow suggests the area of existing Fe-Ge-O ternary compounds.). Crossing this boundary would require highly oxidizing conditions capable of stabilizing Ge and Fe in oxidation states greater than 4+ and 3+, respectively. An intriguing example is Fe$_4$Ge$_2$O$_9$, positioned along the blue line in **Fig. S1**. This compound can be described as a mixture of Fe$^{2+}$/Fe$^{3+}$ with Ge$^{2+}$. The new discovery FeGe$_3$O$_4$ compound sits on the boundary of Fe$^{2+}$ and Ge$^{2+}$.

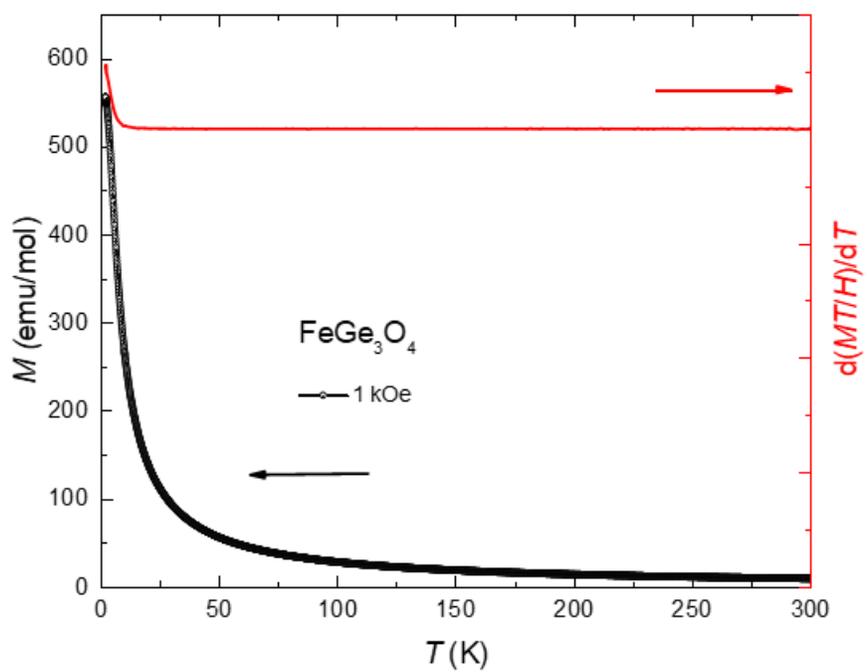

**Fig. S2.** Temperature dependence of magnetic susceptibility at 1000 Oe and d(MT/H)/dT as a function of temperature.

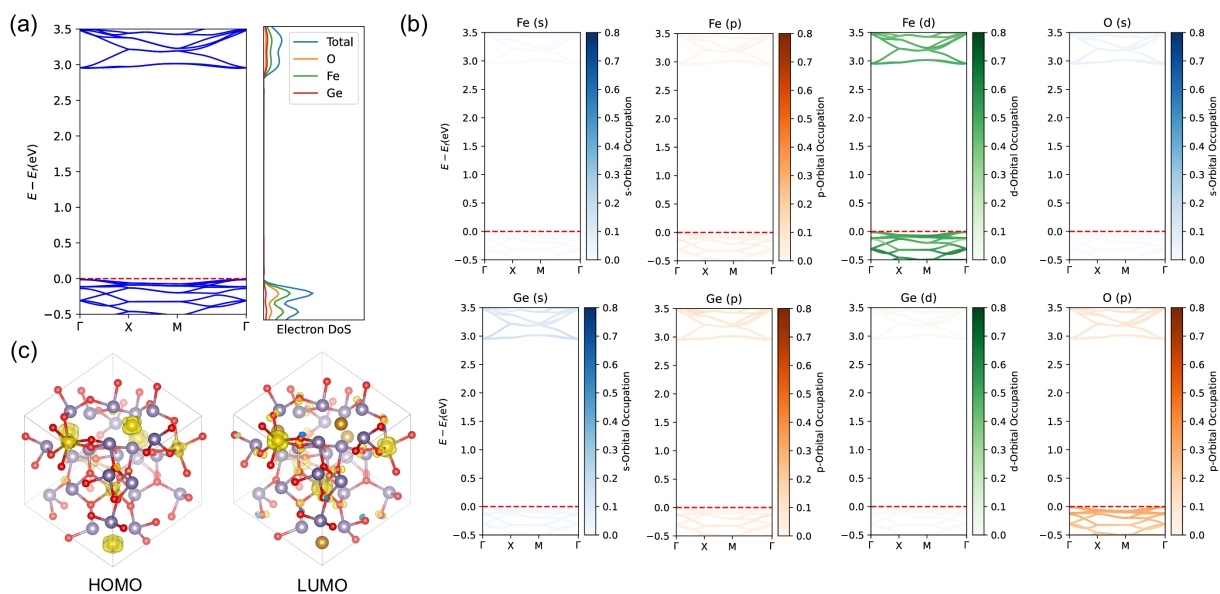

**Fig. S3. Electronic structure of FeGe$_3$O$_4$ calculated by DFT.** (*a*) Electronic band structure and projected density of states (PDOS) of the material, with contributions from O, Fe, and Ge atoms. The Fermi level is set to 0 eV (red dashed line). (*b*) Orbital-projected band structure, showing the contributions of Fe, Ge, and O atoms in different orbitals (s, p, d). The color intensity represents the orbital occupation. (*c*) Partial charge distribution of the highest occupied molecular orbital (HOMO) and the lowest unoccupied molecular orbital (LUMO).

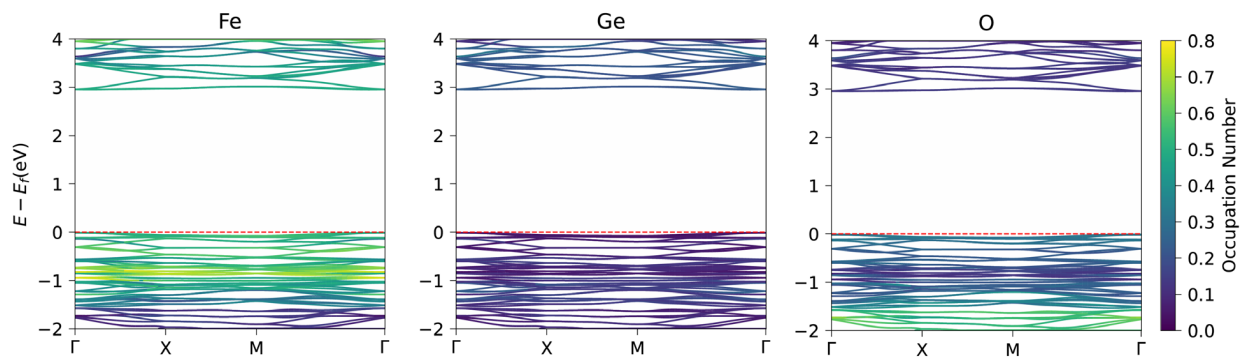

**Fig. S4.** Decomposition of band structure for FeGe$_3$O$_4$ on each of the three elements Fe, Ge and O. The occupation on each band for each element is shown with a colormap from 0 to 1.

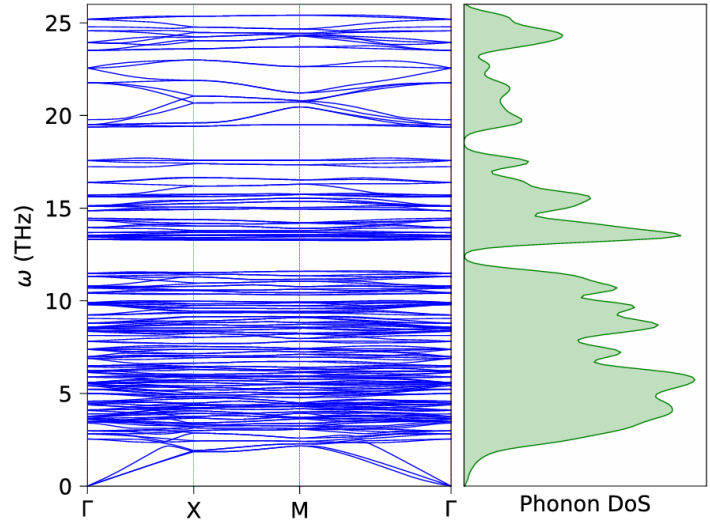

**Fig. S5.** DFT phonon band structure and total density of states for FeGe$_3$O$_4$. The band structure is plotted along the Γ-X-M-Γ symmetry path, and the phonon density of states is smeared out for smoothness.

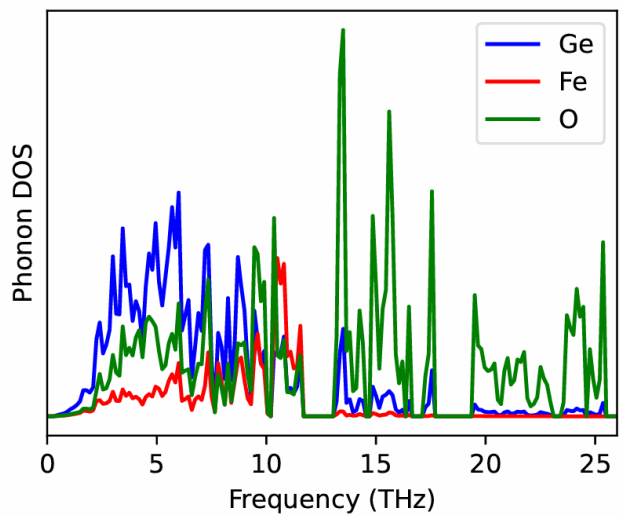

**Fig. S6.** Partial phonon density of states for FeGe$_3$O$_4$, projected on each of the three elements Ge, Fe and O.

To extract the magnetic exchange parameter $J$, we use density functional theory (DFT) to compute the magnetic energies of a series of spin configurations for FeGe$_3$O$_4$. These energies are then mapped onto the Heisenberg spin Hamiltonian with nearest neighbor (NN) couplings $J_1$ and next nearest neighbor (NNN) couplings $J_2$ :

$$H = J_1 \sum_{\langle i,j \rangle} S_i \cdot S_j + J_2 \sum_{\langle\langle i,j \rangle\rangle} S_i \cdot S_j,$$

The DFT results reveal distinct magnetic behavior for the Fe atoms in the system. Among the 8 Fe atoms in the FeGe$_3$O$_4$ unit cell, four of them each surrounded by Ge atoms, exhibit a non-magnetic configuration (with localized spin < 0.04 $\mu_B$) even under spin polarizations, contributing negligibly to the magnetic moment; while the remaining four Fe atoms, located in an $O_h$ symmetry environment coordinated by six O atoms, display a high-spin configuration with magnetic moments exceeding 3.5 $\mu_B$. This is consistent with the MH measurement. Thus, we only consider these four high-spin Fe sites as magnetic, with their spins contributing to the exchange interactions and overall magnetic properties. The calculated results for magnetic exchange energies are tabulated below, where both the nearest-neighbor (NN) coupling indicates ferromagnetic (FM) order in the short length scale. It is worth noticing that as the Hubbard U is turned on to 5.3 eV, where exchange coupling terms get strongly suppressed, the NNN coupling $J_2$ flips sign to a positive number, indicating anti-ferromagnetic (AFM) coupling. Moreover, the AFM coupling $J_2$ is nearly 1/3 of the NN FM coupling $J_1$, and it is very likely that the competition between $J_1$ and $J_2$ leads to the magnetic frustration of FeGe$_3$O$_4$, which leads to the absence of magnetic order on the long range.

**Table S4.** Magnetic exchange energies obtained by DFT energy mapping on the Heisenberg model, with the Hubbard U parameter set as 0 and 5.3 eV. The nearest-neighbor (NN) and next-nearest-neighbor (NNN) coupling correspond to Fe–Fe bond distances of 5.76 Å and 8.61 Å, respectively.

| Ex | Fe–Fe | Bond length (Å) | U = 0 eV | U = 5.3 eV |
|---|---|---|---|---|
| $J_1$ | Fe1–Fe2 | 4.1680(4) | / | / |
| $J_1$ | Fe1–Fe2 | 4.1680(4) | / | / |
| $J_2$ | Fe1–Fe2 | 5.0262(4) | / | / |
| $J_3$ | Fe1–Fe2 | 5.4426(4) | / | / |
| $J_4$ | Fe2–Fe2 | 5.7602(4) | -3.60 K | -0.45 K |
| $J_4$ | Fe1–Fe1 | 5.7694(3) | | |
| $J_5$ | Fe1–Fe1 | 8.4861(4) | -1.61 K | +0.14 K |
| $J_5$ | Fe2–Fe2 | 8.6052(5) | | |

Density Functional Theory (DFT) calculations are conducted using the Vienna Ab initio Simulation Package (VASP). The projector-augmented wave (PAW) method is employed, with exchange-correlation effects described by the Perdew-Burke-Ernzerhof (PBE) formulation of the generalized gradient approximation (GGA). The plane-wave cutoff energy is set at 520 eV to ensure sufficient convergence. The system under investigation, $FeGe_3O_4$, is treated in a non-spin-polarized configuration in consistency with experimental observations. A calibrated Hubbard U correction of 5.3 eV, as used by the Materials Project, is applied to better account for the on-site Coulomb interactions in the Fe $d$ orbitals. Calculations are performed with a 3 × 3 × 3 $k$-point mesh centered at the Gamma point, to sample the Brillouin zone. Geometry optimization is carried out with a convergence criterion for the forces set to 0.02 eV / Å under the symmetric constraint of $P2_13$ space group. Our data analysis on DFT results utilizes the VASPKIT package, and the phonon properties are calculated combining VASP with the Phonopy package with a 2 × 2 × 1 supercell. To understand the magnetic and electronic behaviors of $FeGe_3O_4$, the electronic structure was calculated and shown in **Fig. S3**. The calculated band gap is around 3 eV, indicating the insulating properties of $FeGe_3O_4$. Decomposition of band structure for $FeGe_3O_4$ shown in **Fig. S3b** on each of the three elements Fe, Ge and O shows that the hybridization between Fe and O atoms dominating the Fermi level. [1-7]